\documentclass[12pt,fleqn]{article}
\usepackage{a4wide}
\usepackage{float}
\usepackage{psfig}
\usepackage[dvips]{graphicx}
\usepackage{amssymb}

\def\Rset{\mathbb{R}}

\begin{document}
{\sloppy

\title{{\bf An invariance property of predictors in kernel induced hypothesis spaces}}


\author{Nicola Ancona\thanks{{\normalsize Institute of Intelligent Systems for Automation - C.N.R.
Via Amendola 122/D-I - 70126 Bari - Italy.}}
~\&~
Sebastiano Stramaglia\thanks{{\normalsize TIRES-Center of Innovative Technologies for
Signal Detection and Processing, University of Bari, Italy.}}
\thanks{{\normalsize Dipartimento Interateneo di Fisica, University of Bari, Italy.}}
\thanks{{\normalsize Istituto Nazionale di Fisica Nucleare, Sezione di Bari, Italy.}}
%
}

\date{August 29, 2005}

\maketitle

\begin{abstract}
\noindent {\normalsize We consider kernel based learning methods for regression and
analyze what happens to the risk minimizer when new variables, statistically
independent of input and target variables, are added to the set of input variables;
this problem arises, for example, in the detection of causality relations between two
time series. We find that the risk minimizer remains unchanged if we constrain the risk
minimization to hypothesis spaces induced by suitable kernel functions. We show that
not all kernel induced hypothesis spaces enjoy this property. We present sufficient
conditions ensuring that the risk minimizer does not change, and show that they hold
for inhomogeneous polynomial and Gaussian RBF kernels. We also provide examples of
kernel induced hypothesis spaces whose risk minimizer changes if independent variables
are added as input.}
\end{abstract}


\section{Introduction}
Recent advances in kernel-based learning algorithms have brought
the field of machine learning closer to the goal of autonomy, i.e.
the goal of providing learning systems that require as little
intervention as possible on the part of a human user (Vapnik,
1998). Kernel algorithms work by embedding data into a Hilbert
space, and searching for linear relations in that space. The
embedding is performed implicitly, by specifying the inner product
between pairs of points. Kernel-based approaches are generally
formulated as convex optimization problems, with a single minimum,
and thus do not require heuristic choices of learning rates, start
configuration or other free parameters. On the other hand, the
choice of the kernel and the corresponding feature space are
central choices that generally must be made by a human user. While
this provides opportunities to use prior knowledge about the
problem at hand, in practice it is difficult to find prior
justification for the use of one kernel instead of another
(Shawe-Taylor and Cristianini, 2004). The purpose of this work is
to introduce a novel property enjoyed by some kernel-based
learning machines, which is of particular relevance when a machine
learning approach is developed to evaluate causality between two
simultaneously acquired signals: in this paper we define a
learning machine to be invariant w.r.t.\ independent variables
(property IIV) if it does not change when statistically
independent variables are added to the set of input variables. We
show that the risk minimizer constrained to belong to suitable
kernel induced hypothesis spaces is IIV. This property holds true
for hypothesis spaces induced by inhomogeneous polynomial and
Gaussian kernel functions. We discuss the case of quadratic loss
function and provide sufficient conditions for a kernel machine to
be IIV. We also present examples of kernels which induce spaces
where the risk minimizer is not IIV, and they should not be used
to measure causality.

\section{Preliminaries}
We focus on the problem of predicting the value of a random
variable (r.v.) $s \in \Rset$ with a function $f(\mathbf{x})$ of
the r.v.\ vector $\mathbf{x} \in \Rset^d$. Given a loss function
$V$ and a set of functions called the hypothesis space
$\mathcal{H}$, the best predictor is sought for in $\mathcal{H}$
as the minimizer $f^{*}$ of the \textit{prediction error} or
\textit{generalization error} or \textit{risk} defined as:

\begin{equation}\label{risk-definition-on-x-label}
R \left[ f \right] = \int V \left( s, f(\mathbf{x}) \right)
p(\mathbf{x}, s) d\mathbf{x} ds,
\end{equation}

\noindent where $p(\mathbf{x}, s)$ is the joint density function
of $\mathbf{x}$ and $s$. Given another r.v.\ $\mathbf{y} \in
\Rset^q$, let us add $\mathbf{y}$ to the input variables and
define a new vector appending $\mathbf{x}$ and $\mathbf{y}$, i.e.
$\mathbf{z} = (\mathbf{x}^\top, \mathbf{y}^\top)^\top$. Let us
also consider the predictor $f'^*(\mathbf{z})$ of  $s$, based on
the knowledge of the r.v.\ $\mathbf{x}$ and $\mathbf{y}$,
minimizing the risk:

\begin{equation}\label{risk-definition-on-z-label}
R' \left[ f'\right] = \int V \left( s, f'(\mathbf{z}) \right)
p(\mathbf{z}, s) d\mathbf{z} ds.
\end{equation}

\noindent If $\mathbf{y}$ is statistically independent of
$\mathbf{x}$ and $s$, it is intuitive to require that
$f^*(\mathbf{x})$ and $f'^*(\mathbf{z})$ coincide and have the
same risk. Indeed in this case $\mathbf{y}$ variables do not
convey any information on the problem at hand. The property stated
above is important when predictors are used to identify causal
relations among simultaneously acquired signals, an important
problem with applications in many fields ranging from economy to
physiology (see, e.g., Ancona et al., 2004, and references
therein). The major approach to this problem examines if the
prediction of one series could be improved by incorporating
information of the other, as proposed by Granger, 1969. In
particular, if the prediction error of the first time series is
reduced by including measurements from the second time series in
the regression model, then the second time series is said to have
a causal influence on the first time series. However, not all
prediction schemes are suitable to evaluate causality between two
time series; they should be invariant w.r.t.\ independent
variables, so that, at least asymptotically, they would be able to
recognize variables without causality relationship.

In this work we consider as predictor the function minimizing the risk and we show that
it does not always enjoy this property. In particular we show that if we constrain the
minimization of the risk to suitable hypothesis spaces then the risk minimizer is IIV
(stable under inclusion of independent variables). We limit our analysis to the case of
quadratic loss function $V \left( s, f(\mathbf{x}) \right) = \left( s - f(\mathbf{x})
\right)^2$.

\subsection{Unconstrained $\cal H$}

If we do not constrain the hypothesis space, then $\mathcal{H}$ is
the space of measurable functions for which $R$ is well defined.
It is well known (Papoulis, 1985) that the
minimizer of (\ref{risk-definition-on-x-label}) is the regression
function:
$$
f^*(\mathbf{x}) = \int s p(s | \mathbf{x} ) ds.
$$
\noindent Note that if $\mathbf{y}$ is independent of $\mathbf{x}$
and $s$ then $p(s | \mathbf{x}) = p(s | \mathbf{x}, \mathbf{y})$
and this implies:
$$
f'^*(\mathbf{z}) = \int s p(s| \mathbf{x}, \mathbf{y}) ds = \int s
p(s | \mathbf{x}) ds = f^*(\mathbf{x}).
$$
\noindent Hence the regression function does not change if
$\mathbf{y}$ is also used for predicting $s$; the regression
function is stable under inclusion of independent variables.

\subsection{Linear hypothesis spaces}

Let us consider the case of linear hypothesis spaces:
$$
\mathcal{H} =
\left \{
f | f(\mathbf{x}) =
\mathbf{w}_{\mathbf{x}}^\top \mathbf{x} , \mathbf{w}_{\mathbf{x}}
\in \Rset^d
\right \}.
$$
\noindent Here, and in all the hypothesis spaces that we consider
in this work, we assume that the mean value of the predictor and the
mean of $s$ coincide:
\begin{equation}\label{mean-of-s-and-linear-machine-coincide-label}
E \left \{ s - \mathbf{w}_{\mathbf{x}}^\top \mathbf{x} \right \} = 0,
\end{equation}
\noindent where $E\{ \cdot\}$ means the expectation value. This can be easily achieved by
adding a constant component (equal to one) to the $\mathbf{x}$ vector.
Equation (\ref{mean-of-s-and-linear-machine-coincide-label}) is a sufficient condition for property IIV
in the case of linear kernels. Indeed, let us consider
the risk associated to an element of $\mathcal{H}$:
\begin{equation}\label{risk-definition-on-x-for-linear-predictor-label}
R \left[ \mathbf{w}_{\mathbf{x}} \right] = \int
\left( s - \mathbf{w}_{\mathbf{x}}^\top \mathbf{x} \right)^2
p(\mathbf{x}, s) d\mathbf{x} ds.
\end{equation}
\noindent The parameter vector $\mathbf{w}_{\mathbf{x}}^{*}$,
minimizing the risk, is solution of the following linear system:
\begin{equation}\label{linear-system-obtained-minimizing-the risk-on-x-lin-pred-label}
E \left \{ \mathbf{x} \mathbf{x}^{\top} \right \}
\mathbf{w}_{\mathbf{x}} = E \left \{ s \mathbf{x} \right \}.
\end{equation}
\noindent Let us consider the hypothesis space of linear functions
in the $\mathbf{z} = (\mathbf{x}^\top, \mathbf{y}^\top)^\top$ variable:
$$
\mathcal{H}' =
\left \{
f' | f'(\mathbf{z}) = \mathbf{w}_{\mathbf{z}}^\top \mathbf{z} ,
\mathbf{w}_{\mathbf{z}} \in \Rset^{d+q}
\right \}.
$$
\noindent Writing $\mathbf{w}_{\mathbf{z}} = \left (
\mathbf{w}^\top_{\mathbf{x}}, \mathbf{w}^\top_{\mathbf{y}} \right
)^\top$ with $\mathbf{w}_{\mathbf{y}} \in \Rset^q$, let us
consider the risk associated to an element of $\mathcal{H}'$:
\begin{equation}\label{risk-definition-on-z-for-linear-predictor-label}
R' \left[ \mathbf{w}_{\mathbf{z}} \right] = \int \left( s -
\mathbf{w}_{\mathbf{x}}^\top \mathbf{x} - \mathbf{w}_{\mathbf{y}}^\top \mathbf{y}
\right)^2 p(\mathbf{x}, \mathbf{y}, s) d\mathbf{x} d\mathbf{y} ds.
\end{equation}
\noindent If $\mathbf{y}$ is independent of $\mathbf{x}$ and $s$ then
(\ref{risk-definition-on-z-for-linear-predictor-label}) can be written, due to (\ref{mean-of-s-and-linear-machine-coincide-label}), as:
\begin{equation}\label{risk-definition-on-z-for-linear-predictor-with-Rx-label}
R' \left[ \mathbf{w}_{\mathbf{z}} \right] =
R \left[ \mathbf{w}_{\mathbf{x}} \right] + \int \left(
\mathbf{w}_{\mathbf{y}}^\top \mathbf{y} \right)^2 p(\mathbf{y}) d\mathbf{y}.
\end{equation}
\noindent It follows that the minimum of $R'$ corresponds to
$\mathbf{w}_{\mathbf{y}} = 0$. In conclusion, if $\mathbf{y}$ is
independent of $\mathbf{x}$ and $s$, the predictors $f^*(\mathbf{x})
= \mathbf{w^{*}_{\mathbf{x}}}^\top \mathbf{x}$ and $f'^*(\mathbf{z})
= \mathbf{w^{*}_{\mathbf{z}}}^\top \mathbf{z}$, which minimize the
risks (\ref{risk-definition-on-x-for-linear-predictor-label}) and
(\ref{risk-definition-on-z-for-linear-predictor-label})
respectively, coincide (i.e.,
$f^*(\mathbf{x})=f'^*(\mathbf{x},\mathbf{y})$ for every $\mathbf{x}$
and every $\mathbf{y}$). Moreover the weights associated to the
components of the $\mathbf{y}$ vector are identically null. So the
risk minimizer in linear hypothesis spaces is a IIV predictor.

\section{Nonlinear hypothesis spaces}

Let us now consider nonlinear hypothesis spaces. An important
class of non linear models is obtained mapping the input space to
a higher dimensional feature space and finding a linear predictor
in this new space. Let $\mbox{\boldmath $\phi$}$ be a non linear
mapping function which associates to $\mathbf{x} \in \Rset^{d}$
the vector $\mbox{\boldmath $\phi$}(\mathbf{x}) =
(\phi_1(\mathbf{x}), \phi_2(\mathbf{x}), ...,
\phi_h(\mathbf{x}))^\top \in \Rset^h$, where $\phi_1, \phi_2, ...,
\phi_h$ are $h$ fixed real valued functions. Let us consider
linear predictors in the space spanned by the functions $\phi_i$
for $i=1, 2, ..., h$. The hypothesis space is then:
$$
\mathcal{H} = \left \{
f | f(\mathbf{x}) =
\mathbf{w_{\mathbf{x}}}^\top \mbox{\boldmath $\phi$}(\mathbf{x}),
\mathbf{w_{\mathbf{x}}} \in \Rset^h \right \}.
$$
\noindent In this space, the best linear predictor of $s$ is the function
$f^* \in \mathcal{H}$ minimizing the
risk:
\begin{equation}\label{risk-in-non-linear-h-label}
R \left[ \mathbf{w}_{\mathbf{x}} \right] = \int \left
(s - \mathbf{w_{\mathbf{x}}}^\top \mbox{\boldmath
$\phi$}(\mathbf{x}) \right )^2 p(\mathbf{x}, s) d\mathbf{x} ds.
\end{equation}
\noindent Let us denote $\mathbf{w^*_{\mathbf{x}}}$ the minimizer
of (\ref{risk-in-non-linear-h-label}). We first restrict to the
case of a single additional new feature: let $y$ be a new real
random variable, statistically independent of $s$ and
$\mathbf{x}$, and denote $\gamma^\prime(\mathbf{z})$, with
$\mathbf{z} = (\mathbf{x}^\top, y)^\top$, a generic new feature
involving the $y$ variable. For predicting the r.v.\ $s$ we use
the linear model involving the new feature:
$$
f'(\mathbf{z}) = \mathbf{w_{\mathbf{z}}}^\top
\mbox{\boldmath $\phi$}^\prime ({\mathbf z}),
$$
%
%
\noindent where $\mbox{\boldmath $\phi$}^\prime ({\mathbf z}) =
\left ( \mbox{\boldmath $\phi$} ({\mathbf x})^\top,
\gamma^\prime(\mathbf{z}) \right )^\top$ and
$\mathbf{w_{\mathbf{z}}} = \left ( \mathbf{w_{\mathbf{x}}}^\top, v
\right )^\top$ has to be fixed minimizing:
\begin{equation}\label{risk-in-non-linear-h-on-z-label}
R' \left[ \mathbf{w}_{\mathbf{z}} \right] = \int \left (s -
\mathbf{w_{\mathbf{x}}}^\top \mbox{\boldmath $\phi$}(\mathbf{x}) -
v \gamma^\prime (\mathbf{x},y) \right )^2 p(\mathbf{x}, s)p(y)
d\mathbf{x} dy ds.
\end{equation}
We would like to have $v=0$ at the minimum of $R'$. At
this aim let us evaluate:
$$
\left . {\partial R' \over \partial v} \right |_{0} = - 2 \int
\gamma^\prime (\mathbf{x}, y) \left(s -
\mathbf{w^*_{\mathbf{x}}}^\top \mbox{\boldmath $\phi$}
(\mathbf{x}) \right ) p(\mathbf{x}, s) p(y) d\mathbf{x} dy ds,
$$
where $\partial /\partial|_{0}$ means that the derivative is evaluated at $v = 0$ and
$\mathbf{w_{\mathbf{x}}} = \mathbf{w^*_{\mathbf{x}}}$, where $\mathbf{w^*_{\mathbf{x}}}$
minimizes the risk (\ref{risk-in-non-linear-h-label}). If ${\partial R' /
\partial v}|_{0}$ is not zero, then the predictor is changed after inclusion of feature
$\gamma^\prime$. Therefore ${\partial R' / \partial v}|_{0} =0$ is the
condition that must be satisfied by all the features, involving $y$, to constitute a IIV
({\it stable}) predictor. It is easy to show that if $\gamma^\prime$ does not depend on
${\bf x}$, then this condition holds. More important, it holds if $\gamma^\prime$ is the
product of a function $\gamma (y)$ of $y$ alone and of a component $\phi_i$ of the
feature vector $\mbox{\boldmath $\phi$} (\mathbf{x})$:
\begin{equation}\label{main-property-for-single-feature-label}
\gamma^\prime (\mathbf{x}, y) = \gamma (y) \phi_i (\mathbf{x})
~~~~~ \textrm{for some} ~ i \in \{1,...,h\}.
\end{equation}
Indeed in this case we have:
$$
\left . {\partial R' \over \partial v} \right |_{0} = - 2 \left[
\int \gamma (y) p(y) dy\right] \left[ \int \phi_i (\mathbf{x})
\left ( s - \mathbf{w^*_{\mathbf{x}}}^\top \mbox{\boldmath
$\phi$}(\mathbf{x}) \right ) p(\mathbf{x}, s) d\mathbf{x}
ds\right]=0
$$
because the second integral vanishes as $\mathbf{w^*_{\mathbf{x}}}$ minimizes the risk
(\ref{risk-in-non-linear-h-label}) when only $\mathbf{x}$ variables are used to predict
$s$. We observe that the second derivative
$$
\left . {\partial^2 R' \over \partial v^2} \right |_{0} =  2 \int
\left(\gamma^\prime (\mathbf{x}, y)\right)^2  p(\mathbf{x}, s)
p(y) d\mathbf{x} dy ds
$$
is positive; $(\mathbf{w^*_{\mathbf{x}}},0)$ remains a minimum after inclusion of the $y$ variable. In
conclusion, if the new feature $\gamma^\prime$ involving $y$ verifies
(\ref{main-property-for-single-feature-label}) then the predictor
$f'^*(\mathbf{z})$, which uses both $\mathbf{x}$ and $y$ for predicting $s$,
minimizing (\ref{risk-in-non-linear-h-on-z-label}) and the predictor
$f^*(\mathbf{x})$ minimizing (\ref{risk-in-non-linear-h-label}) coincide.
This shows that the risk minimizer is unchanged after inclusion of $y$ in the input
variables. This preliminary result, which is used in the next subsection,
may be easily seen to hold also for finite-dimensional
vectorial $\mathbf{y}$.

\subsection{Kernel induced hypothesis spaces}

In this section we analyze if our invariance property holds true
in specific hypothesis spaces which are relevant for many learning
schemes such as Support Vector Machines (Vapnik, 1998) and
regularization networks (Evgeniou et al., 2000), just for citing a
few. At this aim in order to predict $s$, we map $\mathbf{x}$ in a
higher dimensional feature space $\mathcal{H}$ by using the
mapping:
$$
\mbox{\boldmath $\phi$} ({\mathbf x}) = (  \sqrt{\alpha_1}\psi_1
({\bf x}), \sqrt{\alpha_2}\psi_2 ({\bf x}), ...,\sqrt{\alpha_h}
\psi_h ({\bf x}), ... ),
$$
where $\alpha_i$ and $\psi_i$  are the eigenvalues and
eigenfunctions of an integral operator whose kernel $K(\mathbf{x},
\mathbf{x}^\prime)$ is a positive definite symmetric function with
the property $K(\mathbf{x}, \mathbf{x}^\prime) = \mbox{\boldmath
$\phi$} ({\mathbf x})^\top \mbox{\boldmath $\phi$} ({\mathbf
x}^\prime)$ (see Mercer's theorem, Vapnik, 1998).

Let us now consider in detail two important kernels.

\subsection{Case $K(\mathbf{x}, \mathbf{x}^\prime) = \left( 1 +
\mathbf{x}^\top \mathbf{x}^\prime \right )^p$}

Let us consider the hypothesis space induced by this kernel:
$$
\mathcal{H} =
\left \{
f | f(\mathbf{x}) = \mathbf{w}_{\mathbf{x}}^\top \mbox{\boldmath
$\phi$} ({\mathbf x}) , \mathbf{w}_{\mathbf{x}} \in
\Rset^{d^\prime}
\right \},
$$
where the components $\phi_i(\mathbf{x})$ of $\mbox{\boldmath
$\phi$} ({\mathbf x})$ are $d^\prime$  monomials, up to $p-th$
degree, which enjoy the following property:
$$
\mbox{\boldmath $\phi$} ({\mathbf x})^\top \mbox{\boldmath $\phi$}
({\mathbf x}^\prime) = (1 + \mathbf{x}^\top \mathbf{x}^\prime)^p.
$$
Let $f^*(\mathbf{x})$ be the minimizer of the risk in
$\mathcal{H}$. Moreover, let $\mathbf{z} =
(\mathbf{x}^\top, \mathbf{y}^\top)^\top$ and consider the hypothesis space
$\mathcal{H}'$ induced by the mapping
$\mbox{\boldmath $\phi^\prime$} ({\mathbf z})$ such that:
$$
\mbox{\boldmath $\phi^\prime$} ({\mathbf z})^\top \mbox{\boldmath
$\phi^\prime$} ({\mathbf z}^\prime) = (1 + \mathbf{z}^\top
\mathbf{z}^\prime)^p.
$$
Let $f'^*(\mathbf{z})$ be the minimizer of the risk in
$\mathcal{H}'$. If $\mathbf{y}$ is independent of $\mathbf{x}$ and
s then $f^*(\mathbf{x})$ and $f'^*(\mathbf{z})$ coincide. In fact
the components of $\mbox{\boldmath $\phi$}^{\prime} ({\mathbf z})$
are all the monomials, in the variables $\mathbf{x}$ and
$\mathbf{y}$, up to the $p-th$ degree: it follows trivially that
$\mbox{\boldmath $\phi$}^{\prime} ({\mathbf z})$ can be written
as:
$$
\mbox{\boldmath $\phi$}^{\prime} ({\mathbf z}) = \left (
\mbox{\boldmath $\phi$} ({\mathbf x})^\top, \mbox{\boldmath
$\gamma$}^{\prime} ({\mathbf z})^\top \right )^\top,
$$
\noindent where each component $\gamma_i^{\prime} ({\mathbf z})$ of the vector
$\mbox{\boldmath $\gamma$}^{\prime} ({\mathbf z})$ verifies
(\ref{main-property-for-single-feature-label}), that is it is given by the product of a
component $\phi_j ({\mathbf x})$ of the vector $\mbox{\boldmath $\phi$} ({\mathbf x})$
and of a function $\gamma_i ({\mathbf y})$ of the variable $\mathbf{y}$ only:
$$
\gamma_i^{\prime} ({\mathbf z}) = \phi_j ({\mathbf x}) \gamma_i ({\mathbf y}).
$$
As an example, we show this property for the case of $\mathbf{x} =
(x_1, x_2)^\top$, $\mathbf{z} = (x_1, x_2, y)^\top$ and $p = 2$.
In this case the mapping functions $\mbox{\boldmath $\phi$}
({\mathbf x})$ and $\mbox{\boldmath $\phi$}^{\prime} ({\mathbf
z})$ are:
$$
\mbox{\boldmath $\phi$} ({\mathbf x}) = (1, \sqrt{2}x_1,
\sqrt{2}x_2, \sqrt{2} x_1 x_2, x_1^2, x_2^2)^\top,
$$
$$
\mbox{\boldmath $\phi$}^\prime ({\mathbf z}) = (1, \sqrt{2}x_1,
\sqrt{2}x_2, \sqrt{2} x_1 x_2, x_1^2, x_2^2, \sqrt{2} y, \sqrt{2}
x_1 y, \sqrt{2} x_2 y, y^2)^\top,
$$
where one can easily check that $\mbox{\boldmath $\phi$} ({\mathbf
x})^\top \mbox{\boldmath $\phi$} ({\mathbf x}^\prime) = (1 +
\mathbf{x}^\top \mathbf{x}^\prime)^2$ and $\mbox{\boldmath
$\phi$}^\prime ({\mathbf z})^\top \mbox{\boldmath $\phi$}^\prime
({\mathbf z}^\prime) = (1 + \mathbf{z}^\top \mathbf{z}^\prime)^2$.
In this case the vector $\mbox{\boldmath $\gamma^{\prime}$}
({\mathbf z})$ is:
$$
\mbox{\boldmath $\gamma^{\prime}$} ({\mathbf z}) = \left (
\phi_1(\mathbf{x}) \sqrt{2} y, \phi_2(\mathbf{x}) y,
\phi_3(\mathbf{x}) y, \phi_1(\mathbf{x}) y^2 \right )^\top.
$$
According to the argument described before, the risk minimizer in this hypothesis space
satisfies the invariance property.

Note that, remarkably, the risk minimizer in the hypothesis space induced by the homogeneous polynomial kernel
$K(\mathbf{x}, \mathbf{x}^\prime) = \left( \mathbf{x}^\top \mathbf{x}^\prime \right )^p$
does not have the invariance property for a generic probability density, as one can
easily check working out explicitly the $p=2$ case.

\subsection{Translation invariant kernels}

In this section we present a formalism which generalizes our
discussion to the case of hypothesis spaces whose features
constitute an uncountable set. We show that the IIV property holds
for linear predictors on feature spaces induced by translation
invariant kernels. In fact let $K(\mathbf{x}, \mathbf{x}^\prime) =
K(\mathbf{x} - \mathbf{x}^\prime)$ be a positive definite kernel
function, with $\mathbf{x}, \mathbf{x}^\prime \in \Rset^d$. Let
$\tilde{K}(\mbox{\boldmath $\omega_{x}$})$ be the Fourier
transform of $K(\mathbf{x})$: $K(\mathbf{x}) \leftrightarrow
\tilde{K}(\mbox{\boldmath $\omega_{x}$})$. For the time shifting
property we have that: $K(\mathbf{x} - \mathbf{x}^\prime)
\leftrightarrow \tilde{K}(\mbox{\boldmath $\omega_{x}$}) e^{-j}
\mbox{\boldmath $^{\omega_x^\top}$} \mathbf{^{x^\prime}}$. By
definition of the inverse Fourier transform, neglecting constant
factors, we know that (Girosi, 1998):
$$
K(\mathbf{x} - \mathbf{x}^\prime) = \int_{\Rset^d}
\tilde{K}(\mbox{\boldmath $\omega_{x}$}) e^{-j} \mbox{\boldmath
$^{\omega_{x}^\top}$} \mathbf{^{x^\prime}} e^{j} \mbox{\boldmath
$^{\omega_{x}^\top}$} \mathbf{^{x}} d \mbox{\boldmath
$\omega_{x}$}.
$$
\noindent Being $K$ positive definite we can write:
$$
K(\mathbf{x} - \mathbf{x}^\prime) = \int_{\Rset^d} \sqrt{\tilde{K}
(\mbox{\boldmath $\omega_{x}$})} e^{j} \mbox{\boldmath
$^{\omega_{x}^\top}$} \mathbf{^{x}} \left ( \sqrt{\tilde{K}
(\mbox{\boldmath $\omega_{x}$})} e^{j} \mbox{\boldmath
$^{\omega_{x}^\top}$} \mathbf{^{x^\prime}} \right )^* d
\mbox{\boldmath $\omega_{x}$},
$$
\noindent where $*$ indicates conjugate. Then we can write
$K(\mathbf{x}, \mathbf{x}^\prime) = \left \langle
\phi{\mbox{\boldmath $_x$}}, \phi{\mbox{\boldmath $_{x^\prime}$}}
\right \rangle$ where:
\begin{equation}\label{phi-sub-x-of-omega-label}
\phi{\mbox{\boldmath $_x$}} (\mbox{\boldmath $\omega_{x}$}) =
\sqrt{\tilde{K}(\mbox{\boldmath $\omega_{x}$})} e^{j}
\mbox{\boldmath $^{\omega_{x}^\top}$} \mathbf{^{x}}
\end{equation}
\noindent are the generalized eigenfunctions. Note that, in this
case, the mapping function $\phi{\mbox{\boldmath $_x$}}$
associates a function to $\mathbf{x}$, that is
$\phi{\mbox{\boldmath $_x$}}$ maps the input vector $\mathbf{x}$
in a feature space with an infinite and uncountable number of
features. Let us consider the hypothesis space induced by $K$:
$$
\mathcal{H} =
\left \{
f | f(\mathbf{x}) = \left \langle w{\mbox{\boldmath $_x$}} ,
\phi{\mbox{\boldmath $_x$}} \right \rangle , w{\mbox{\boldmath
$_x$}} \in {\cal W}{\mbox{\boldmath $_x$}}
\right \},
$$
where:
\begin{equation}\label{complex-scalar-product-label}
\left \langle w{\mbox{\boldmath $_x$}} , \phi{\mbox{\boldmath
$_x$}} \right \rangle = \int_{\Rset^d} w{\mbox{\boldmath $_x$}}
(\mbox{\boldmath $\omega_{x}$}) \phi{\mbox{\boldmath $_x^*$}}
(\mbox{\boldmath $\omega_{x}$}) d \mbox{\boldmath $\omega_{x}$},
\end{equation}
\noindent and ${\cal W}{\mbox{\boldmath $_x$}}$ is the set of complex measurable
functions for which (\ref{complex-scalar-product-label}) is well defined and
real\footnote{{\normalsize In particular elements of ${\cal W}{\mbox{\boldmath $_x$}}$
satisfy $w{\mbox{\boldmath $_x$}}(-\mbox{\boldmath $\omega_{x}$})=w{\mbox{\boldmath
$_x^*$}}(\mbox{\boldmath $\omega_{x}$})$.}}. Note that $w{\mbox{\boldmath $_x$}}$ is
now a complex function, it is not a vector anymore. In this space the best linear
predictor is the function $\bar{f}= \left \langle \bar{w}{\mbox{\boldmath $_x$}} ,
\phi{\mbox{\boldmath $_x$}} \right \rangle$ in $\mathcal{H}$ minimizing the risk
functional:
$$
R \left [ w{\mbox{\boldmath $_x$}} \right ] = E \left \{
\left ( s - \left \langle w{\mbox{\boldmath $_x$}} ,
\phi{\mbox{\boldmath $_x$}} \right \rangle \right )^2
\right \}
$$
It is easy to show that the optimal function
$\bar{w}{\mbox{\boldmath $_x$}}$ is solution of the following
integral equation:
\begin{equation}\label{integral-equation-on-x-label}
E \left \{
s e^{-j} \mbox{\boldmath $^{\omega_{x}^\top}$} \mathbf{^{x}}
\right \} =
\int_{\Rset^d} w{\mbox{\boldmath $_x$}} (\mbox{\boldmath
$\xi_{x}$}) \sqrt{\tilde{K}(\mbox{\boldmath $\xi_{x}$})}
\Phi{\mbox{\boldmath $_x^*$}}(\mbox{\boldmath $\omega_{x}$} +
\mbox{\boldmath $\xi_{x}$}) d \mbox{\boldmath $\xi_{x}$},
\end{equation}
where $\mbox{\boldmath $\xi_{x}$}$ is a dummy integration variable and
$\Phi{\mbox{\boldmath $_x$}}(\mbox{\boldmath $\omega_{x}$}) = E \left\{ e^{j}
\mbox{\boldmath $^{\omega_{x}^\top}$} \mathbf{^{x}} \right\}$ is the characteristic
function\footnote{{\normalsize $\Phi{\mbox{\boldmath $_x$}}(-\mbox{\boldmath
$\omega_{x}$})$ is the Fourier transform of the probability density $p(\mathbf{x})$ of
the r.v.\ $\mathbf{x}$.}} of the r.v.\ $\mathbf{x}$ (Papoulis, 1985).
%
%
Let us indicate $\tilde{F}(\mbox{\boldmath $\omega_{x}$}) =
w{\mbox{\boldmath $_x$}} (\mbox{\boldmath $\omega_{x}$})
\sqrt{\tilde{K}(\mbox{\boldmath $\omega_{x}$})}$ and
$\tilde{G}(\mbox{\boldmath $\omega_{x}$}) = E \left \{
s e^{j} \mbox{\boldmath $^{\omega_{x}^\top}$} \mathbf{^{x}}
\right \}$. Then (\ref{integral-equation-on-x-label}) can be
written as:
$$
\tilde{G}(\mbox{\boldmath $\omega_{x}$}) =
\tilde{F}(\mbox{\boldmath $\omega_{x}$}) \star
\Phi{\mbox{\boldmath $_x$}}(\mbox{\boldmath $\omega_{x}$}),
$$
\noindent where $\star$ indicates cross-correlation between
complex functions. In the spatial domain this implies:
$$
G(\mathbf{x}) = F^{*}(\mathbf{x}) p(-\mathbf{x}).
$$
\noindent In conclusion, assuming that the density $p(\mathbf{x})$
is strictly positive, the function $\bar{w}{\mbox{\boldmath $_x$}}
(\mbox{\boldmath $\omega_{x}$})$ minimizing the risk is unique and
it is given by:
$$
\bar{w}{\mbox{\boldmath $_x$}} (\mbox{\boldmath $\omega_{x}$}) =
\mathcal{F} \left \{ G^{*}(\mathbf{x}) / p(-\mathbf{x}) \right \}
/ \sqrt{\tilde{K}(\mbox{\boldmath $\omega_{x}$})}.
$$
Substituting this expression into equation
(\ref{complex-scalar-product-label}) leads to
$$
\bar{f}(\mathbf{x}) = \int s p(s | \mathbf{x} ) ds,
$$
i.e. the risk minimizer coincides with the regression function. In
other words, the hypothesis space $\mathcal{H}$, induced by $K$,
is sufficiently large to contain the regression function. This
proves that translation invariant kernels are IIV.

It is interesting to work out and explicitly prove the IIV
property in the case of translation invariant and separable
kernels. As in the previous section, let $\mathbf{y} \in \Rset^q$
be a r.v.\ vector \emph{independent} of $\mathbf{x}$ and $s$ and
use the vector $\mathbf{z} = (\mathbf{x}^\top,
\mathbf{y}^\top)^\top$ for predicting $s$. At this aim, let us
consider the following mapping function:
\begin{equation}\label{phi-sub-z-of-omega-z-label}
\phi{\mbox{\boldmath $_z^\prime$}} (\mbox{\boldmath $\omega_{z}$})
= \sqrt{\tilde{K}^{\prime}(\mbox{\boldmath $\omega_{z}$})} e^{j}
\mbox{\boldmath $^{\omega_{z}^\top}$} \mathbf{^{z}},
\end{equation}
\noindent where $\mbox{\boldmath $\omega_{z}$} = (\mbox{\boldmath
$\omega_{x}$}^\top, \mbox{\boldmath $\omega_{y}$}^\top)^{\top}$
and $\left \langle \phi{\mbox{\boldmath $_z^\prime$}} ,
\phi{\mbox{\boldmath $_{z^\prime}^\prime$}} \right \rangle =
K^{\prime} (\mathbf{z} - \mathbf{z}^{\prime})$. Let us consider
the hypothesis space induced by $K^{\prime}$:
$$
\mathcal{H}' =
\left \{
f'| f'(\mathbf{z}) = \left \langle w{\mbox{\boldmath $_z$}} ,
\phi{\mbox{\boldmath $_z^\prime$}} \right \rangle ,
w{\mbox{\boldmath $_z$}} \in {\cal W}{\mbox{\boldmath $_z$}}
\right \}.
$$
The best linear predictor is the function $\bar{f}' =
\left \langle \bar{w}_{\mathbf{z}} , \phi^{\prime}_{\mathbf{z}}
\right \rangle$ in $\mathcal{H}'$
minimizing the risk functional:
$$
R' \left [ w{\mbox{\boldmath $_z$}} \right ] = E \left \{
\left ( s - \left \langle w{\mbox{\boldmath $_z$}} ,
\phi{\mbox{\boldmath $_z^\prime$}} \right \rangle \right )^2
\right \},
$$
where the optimal function $\bar{w}{\mbox{\boldmath $_z$}}$ is
solution of the integral equation (see
(\ref{integral-equation-on-x-label})):
\begin{equation}\label{integral-equation-on-z-label}
E \left \{
s e^{-j} \mbox{\boldmath $^{\omega_{z}^\top}$} \mathbf{^{z}}
\right \} =
\int_{\Rset^{d+q}} w{\mbox{\boldmath $_z$}} (\mbox{\boldmath
$\xi_{z}$}) \sqrt{\tilde{K}^{\prime}(\mbox{\boldmath $\xi_{z}$})}
\Phi{\mbox{\boldmath $_z^*$}}(\mbox{\boldmath $\omega_{z}$} +
\mbox{\boldmath $\xi_{z}$}) d \mbox{\boldmath $\xi_{z}$},
\end{equation}
\noindent where $\mbox{\boldmath $\omega_{z}$} = \left (
\mbox{\boldmath $\omega_{x}$}^\top, \mbox{\boldmath
$\omega_{y}$}^\top \right )^\top$. Note that, being $\mathbf{x}$
and $\mathbf{y}$ independent, the characteristic function of
$\mathbf{z}$ factorizes:
$$
\Phi{\mbox{\boldmath $_z$}}(\mbox{\boldmath $\omega_{z}$})=
\Phi{\mbox{\boldmath $_x$}}(\mbox{\boldmath
$\omega_{x}$})\Phi{\mbox{\boldmath $_y$}}(\mbox{\boldmath
$\omega_{y}$}).
$$
If $K^{\prime}(\mathbf{z})$ is separable:
\begin{equation}\label{k-separable-kernel-label}
K^{\prime}(\mathbf{z}) = K(\mathbf{x})H(\mathbf{y})
\end{equation}
then its Fourier transform takes the form of $\tilde{K}^{\prime}
(\mbox{\boldmath $\omega_{z}$}) = \tilde{K} (\mbox{\boldmath
$\omega_{x}$}) \tilde{H} (\mbox{\boldmath $\omega_{y}$})$. Being
$
E \left \{
s e^{-j} \mbox{\boldmath $^{\omega_{z}^\top}$} \mathbf{^{z}}
\right \} = E \left \{
s e^{-j} \mbox{\boldmath $^{\omega_{x}^\top}$} \mathbf{^{x}}
\right \} E \left \{
e^{-j} \mbox{\boldmath $^{\omega_{y}^\top}$} \mathbf{^{y}}
\right \} $ , (\ref{integral-equation-on-z-label}) becomes:
\begin{eqnarray}\label{integral-equation-on-z-with-k-separable-label}
\nonumber \lefteqn{
E \left \{
s e^{-j} \mbox{\boldmath $^{\omega_{x}^\top}$} \mathbf{^{x}}
\right \}
\Phi{\mbox{\boldmath $_y^*$}}(\mbox{\boldmath $\omega_{y}$})
=} \hspace{2.0cm} \\
& &
\int_{\Rset^{d+q}} w{\mbox{\boldmath $_z$}} (\mbox{\boldmath
$\xi_{z}$})
\sqrt{\tilde{K}(\mbox{\boldmath $\xi_{x}$})}
\sqrt{\tilde{H}(\mbox{\boldmath $\xi_{y}$})}
\Phi{\mbox{\boldmath $_x^*$}}(\mbox{\boldmath $\omega_{x}$} +
\mbox{\boldmath $\xi_{x}$})
\Phi{\mbox{\boldmath $_y^*$}}(\mbox{\boldmath $\omega_{y}$} +
\mbox{\boldmath $\xi_{y}$})
d \mbox{\boldmath $\xi_{z}$}.
\end{eqnarray}
\noindent
The risk minimizer $\bar{w}_{\mathbf{z}}$ solution of
(\ref{integral-equation-on-z-with-k-separable-label}) is:
\begin{equation}\label{w-separable-label}
\bar{w}{\mbox{\boldmath $_z$}} (\mbox{\boldmath $\omega_{x}$},
\mbox{\boldmath $\omega_{y}$}) = \bar{w}{\mbox{\boldmath $_x$}}
(\mbox{\boldmath $\omega_{x}$}) {\delta (\mbox{\boldmath
$\omega_{y}$})\over \sqrt{\tilde{H}(0)}}.
\end{equation}
\noindent This can be checked substituting
(\ref{w-separable-label}) in
(\ref{integral-equation-on-z-with-k-separable-label}) and using
equation (\ref{integral-equation-on-x-label}). The structure of
eq.(\ref{w-separable-label}) guarantees that the predictor is
unchanged under inclusion of variables $\mathbf{y}$. This is the
case, in particular, for the Gaussian RBF kernel. Finally note
that a property similar to
(\ref{main-property-for-single-feature-label}) holds true also in
this hypothesis space. In fact, as $K^{\prime}$ is separable,
(\ref{phi-sub-z-of-omega-z-label}) implies that:
\begin{equation}\label{property-of-phi-sub-z-of-omega-z-label}
\phi{\mbox{\boldmath $_z^\prime$}} (\mbox{\boldmath $\omega_{z}$})
= \phi{\mbox{\boldmath $_x$}} (\mbox{\boldmath $\omega_{x}$})
\gamma{\mbox{\boldmath $_y$}} (\mbox{\boldmath $\omega_{y}$})
\end{equation}
\noindent where $\gamma{\mbox{\boldmath $_y$}} (\mbox{\boldmath
$\omega_{y}$}) =
\sqrt{\tilde{H}(\mbox{\boldmath $\omega_{y}$})} e^{j}
\mbox{\boldmath $^{\omega_{y}^\top}$} \mathbf{^{y}}$
with the property:
$\left\langle \gamma{\mbox{\boldmath $_y$}} ,
\gamma{\mbox{\boldmath $_{y^\prime}$}} \right\rangle =
H(\mathbf{y} - \mathbf{y}^{\prime})$. Eq. (\ref{property-of-phi-sub-z-of-omega-z-label})
may be seen as a {\it continuum} version of property
(\ref{main-property-for-single-feature-label}).

\section{Discussion}
In this work we  consider, in the frame of kernel methods for
regression, the following question: does the risk minimizer change
when statistically independent variables are added to the set of
input variables? We show that this property is guaranteed by not all
the hypothesis spaces. We outline sufficient conditions ensuring
this property, and show that it holds for inhomogeneous polynomial
and Gaussian RBF kernels. Whilst these results are relevant to
construct machine learning approaches to study causality between
time series, in our opinion they might also be important in the more
general task of kernel selection. Our discussion concerns the risk
minimizer, hence it holds only in the asymptotic regime; the
analysis of the practical implications of our results, i.e. when
only a finite data set is available to train the learning machine,
is matter for further research. It is worth noting, however, that
our results hold also for a finite set of data, if the probability
distribution is replaced by the empirical measure. Another
interesting question is how this scenario changes when a
regularization constraint is imposed on the risk minimizer (Poggio
et al., 1990) and loss functions different from the quadratic one
are considered. Moreover it would be interesting to analyze the
connections between our results and classical problems of machine
learning such as feature selection and sparse representation, that
is the determination of a solution with only a few number of non
vanishing components. If we look for the solution in overcomplete or
redundant spaces of vectors or functions, where more than one
representation exists, then it makes sense to impose a sparsity
constraint on the solution. In the case here considered, the
sparsity of $\mathbf{w^{*}}$ emerges as a consequence of the
existence of independent input variables using a quadratic loss
function.

The authors thank two anonymous reviewers whose comments were
valuable to improve the presentation of this work.
%
%

\noindent {\bf References}

\noindent {\bf Ancona, N., Marinazzo, D., and Stramaglia, S.}
(2004). Radial basis function approach to nonlinear Granger
causality of time series. Physical Review E, 70, 56221-56227.

\noindent {\bf Evgeniou, T., Pontil, M., and Poggio, T.} (2000).
Regularization Networks and Support Vector Machines. Advances in
Computational Mathematics, 13(1), 1-50.

\noindent {\bf Girosi, F. }(1998) An equivalence between sparse
approximation and Support Vector Machines. Neural Computation, 10,
1455-1480.

\noindent {\bf Granger, C.W.J.} (1969). Testing for causality and
feedback. Econometrica, 37, 424-438.

\noindent {\bf Papoulis, A.} (1985). {\it Probability, Random
Variables, and Stochastic Processes}. McGraw-Hill International
Student Edition.

\noindent {\bf Poggio, T., and Girosi, F.} (1990). Regularization
algorithms for learning that are equivalent to multilayer networks.
Science, 247, 978-986.

\noindent {\bf Shawe-Taylor, J., and Cristianini, N.} (2004). {\it
Kernel Methods For Pattern Analysis}. Cambridge University Press.

\noindent {\bf Vapnik, V.} (1998). {\it Statistical Learning
Theory}. John Wiley \& Sons, INC.
%

\end{document}